\documentclass[runningheads]{llncs}
\usepackage{graphicx}
\usepackage{float}
\usepackage[table]{xcolor}
\usepackage{enumitem}
\usepackage{balance}
\usepackage{hyperref}
\usepackage{xcolor}
\usepackage{fontawesome5}
\usepackage{comment}
\usepackage{subcaption}
\usepackage{tabularx}

\usepackage{tikz}

\hypersetup{
  colorlinks   = True, %Colours links instead of ugly boxes
  urlcolor     = blue, %Colour for external hyperlinks
  linkcolor    = blue, %Colour of internal links
  citecolor    = blue %Colour of citations
}

\begin{document}

\title{A Decision Model for Selecting Patterns and Strategies to Decompose Applications into Microservices}

\titlerunning{Decision Model for Microservices Decomposition}
\author{Muhammad Waseem$^{1}$, Peng Liang$^{1*}$, Gast\'{o}n M\'{a}rquez$^{2}$\\Mojtaba Shahin$^{3}$, Arif Ali Khan$^{4}$, Aakash Ahmad$^{5}$}

\author{Muhammad Waseem\inst{1}\and Peng Liang\inst{1*}\and Gast\'{o}n M\'{a}rquez\inst{2}\\ Mojtaba Shahin\inst{3}\and Arif Ali Khan\inst{4}\and Aakash Ahmad\inst{5}}
\authorrunning{M. Waseem et al.}
\institute{School of Computer Science, Wuhan University, Wuhan, China\\ 
\email{\{m.waseem,liangp\}@whu.edu.cn}\\
\and Department of Electronics and Informatics, Federico Santa María Technical University, Concepci\'{o}n, Chile\\
\and Faculty of Information Technology, Monash University, Melbourne, Australia\\
\and Faculty of Information Technology, University of Jyvaskyla, Jyvaskyla, Finland\\
\and College of Computer Science and Engineering, University of Ha'il, Ha'il, Saudi Arabia
}

\maketitle              
\begin{abstract}
Microservices Architecture (MSA) style is a promising design approach to develop software applications consisting of multiple small and independently deployable services. Over the past few years, researchers and practitioners have proposed many MSA patterns and strategies covering various aspects of microservices design, such as application decomposition. However, selecting appropriate patterns and strategies can entail various challenges for practitioners. To this end, this study proposes a decision model for selecting patterns and strategies to decompose applications into microservices. We used peer-reviewed and grey literature to collect the patterns, strategies, and quality attributes for creating this decision model.
\end{abstract}

\section{Introduction}
\label{sec:introduction}

%microservices
Microservices Architecture (MSA) inspired by Service-Oriented Architecture (SOA) has gained immense popularity in the past few years \cite{dragoni2017microservices}. With MSA, an application is designed as a set of business-driven microservices that can be developed, deployed, tested, and scaled independently \cite{taibi2019monolithic}. Organizations adopt MSA due to better availability, scalability, productivity, performance, fault-tolerance, and cloud support compared with SOA or monolithic applications. It is argued that MSA can also help build autonomous development teams \cite{taibi2019monolithic}. %\cite{taibi2017processes}.

Microservices systems entail a significant degree of complexity at the design phase and runtime configurations from an architecture perspective \cite{newman2020building}. Haselbock et al. \cite{AR5} identified several design areas for microservices systems, such as application decomposition, microservices security, and communication. On the other hand, literature reviews (e.g.,~\cite{waseemMSAdevops}), existing practices (e.g.,~\cite{waseemMSAdesign}), and exploratory studies (e.g.,~\cite{waseem2021nature}) indicate several challenges related to the design areas mentioned in \cite{AR5}, for instance, clearly defining the boundaries of microservices, addressing their security concerns, and managing the communication between a large number of microservices.

Both academia and industry have presented reusable solutions for microservices systems in the form of patterns and strategies. These patterns and strategies are currently distributed across different publications (e.g., scientific and grey literature). The practitioners need to navigate pattern to pattern (and strategy) until a suitable combination of patterns (and strategies) that can address the challenges is found. Moreover, the practitioners cannot find a holistic view of the patterns and strategies for a trade-off analysis (e.g., patterns influence Quality Attributes (QAs)). According to the recent studies (e.g., \cite{waseemMSAdevops}\cite{waseem2021nature}\cite{waseemMSAdesign}), most of the design, development, monitoring and testing challenges are raised when application is decomposed into microservices. To this end, we present a \textbf{decision model} that assists practitioners in selecting appropriate patterns and strategies for decomposing applications into microservices. Decision models are a structured way of exploring the problem and solution space to achieve the design goal \cite{AR7}. The proposed decision model has been developed based on a mini multivocal literature review through reviewing the scientific and grey literature.

\textbf{Paper Organization}: Section \ref{ModelingModels} describes decision models and modeling nations; Section \ref{sec:modelsdescripation} presents the details of the application decomposition decision model; Section \ref{sec:threats} discusses the threats to validity; Section \ref{sec:relatedWork} presents related work; Section \ref{sec:conclusions} concludes this work with future research directions.

\section{Modeling Decision Model}

\label{ModelingModels}
The decision models in software architecture are used to map elements of the problem space to elements of the solution space \cite{AR7}. The problem space represents functional and non-functional requirements, whereas the solution space represents design elements \cite{AR7}. To create decision models for microservices systems, we represent the problem space as a set of QAs and the solution space as a set of microservices patterns and strategies. We developed the decision model for application decomposition because most of the design, development, monitoring, testing, and deployment challenges in microservices systems are rooted in this area (\cite{waseemMSAdevops}\cite{waseem2021nature}\cite{waseemMSAdesign}).  
We collected required patterns, strategies, QAs, and impact of patterns on QAs for creating the decision model based on a mini multivocal literature review.

\begin{figure}[!t]
 \centering
\includegraphics[width=0.75\textwidth]{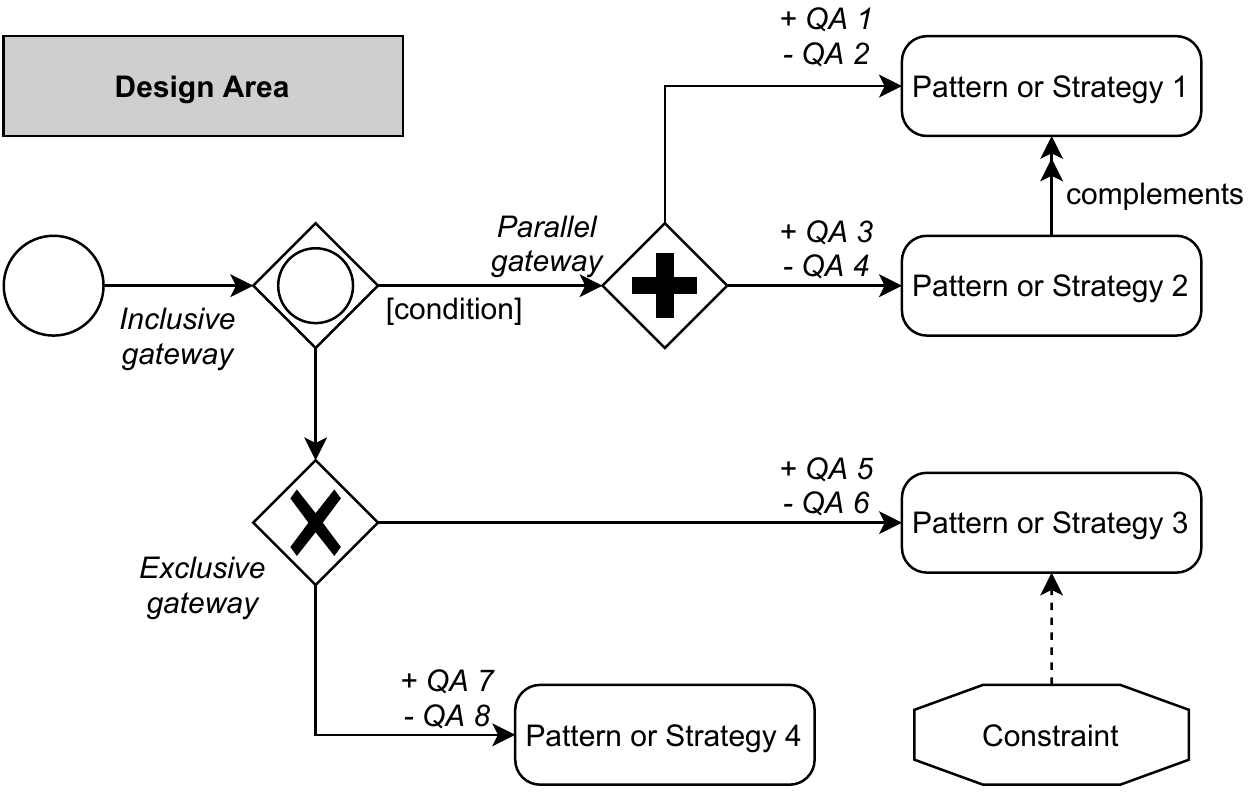}
\caption{Notations used in the decision models}
\label{fig:researchThemeProcess}
\end{figure}

Fig. \ref{fig:researchThemeProcess} presents the notations used in the decision model presented in this paper. We used \textit{Inclusive}, \textit{Exclusive}, and \textit{Parallel} gateways of Business Process Model and Notation (BPMN) for indicating the decision flow. An MSA design area is represented through \textit{grey box}. A \textit{circle} is used to denote the start of a decision process. An \textit{Inclusive gateway} is used to trigger more than one outgoing paths within a decision process. An \textit{Exclusive gateway} is used to trigger one of the outgoing paths. In contrast, A \textit{Parallel gateway} represents the multiple parallel outgoing paths within a decision process. We used \textit{rounded rectangle} to represent the patterns and strategies belong to an MSA design area. A \textit{double-headed} arrow shows a “complements” relationship between two patterns or strategies. An \textit{octagon and dashed} arrow is used to represent the constraints of each pattern or strategy. Finally, plus (+) and minus (-) signs indicate the positive and negative impact of each pattern or strategy on the QAs.

\section{Application Decomposition Decision Model}
\label{sec:modelsdescripation}

\begin{figure*}[!t]
 \centering
\includegraphics[width=0.75\textwidth]{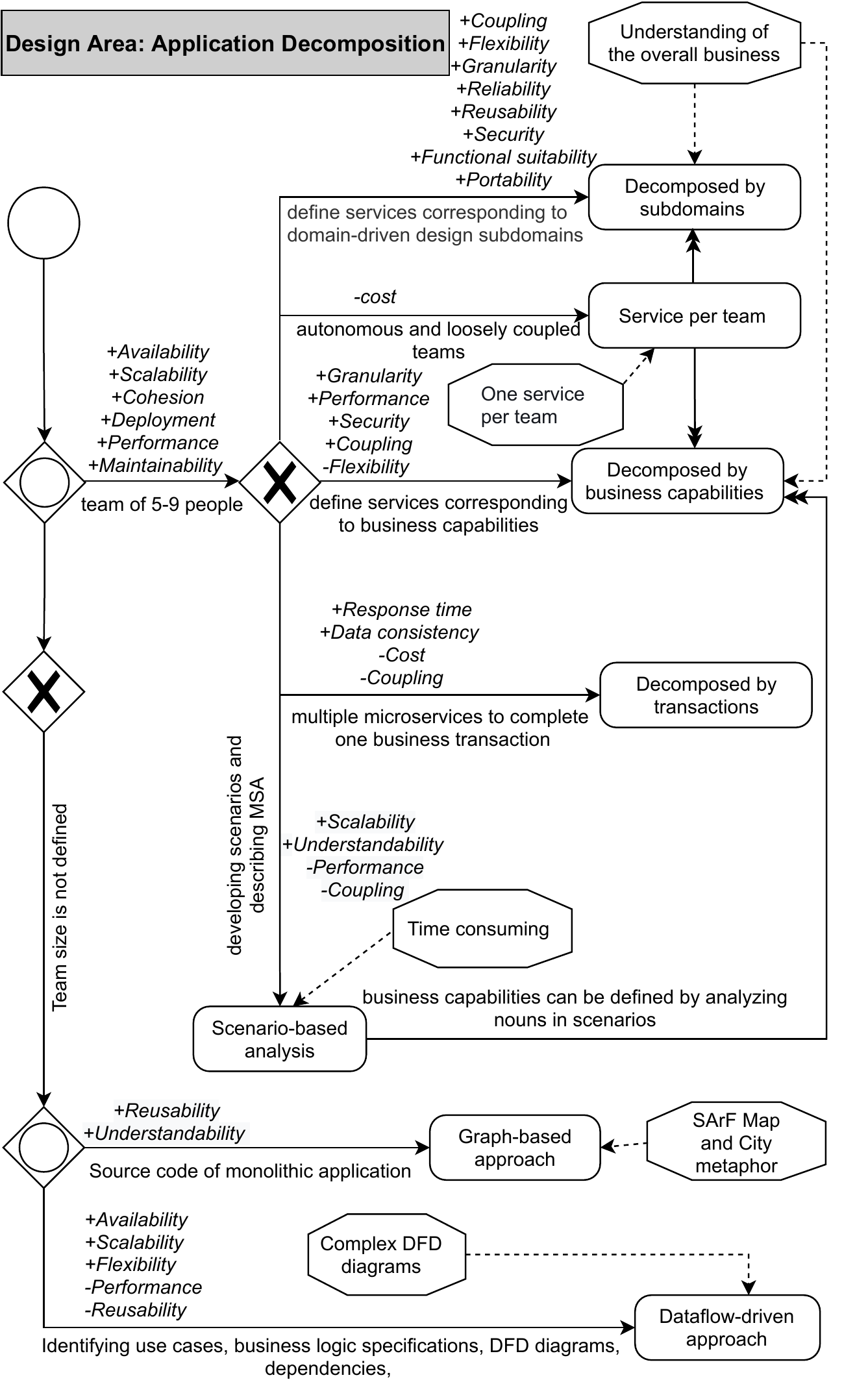}
\caption{Decision model for application decomposition}
\label{fig:AppDecopmositionModel}
\end{figure*}

Monolithic applications need to be decomposed into small, independent, and loosely coupled microservices to achieve the benefits (e.g., improved scalability, independent deployment). There are several ways to break down an application into microservices. The patterns and strategies collected are used to create the application decomposition decision model (see Fig. \ref{fig:AppDecopmositionModel}). The decision process for application decomposition into microservices is based on the team size and impact of patterns and strategies on QAs. If the application needs to be decomposed into microservices for the team of 5-9 people to increase \textit{Availability}, \textit{Scalability}, \textit{Cohesion}, \textit{Deployment}, \textit{Performance}, and \textit{Maintainability}, we can use one among five illustrated patterns (see Fig. \ref{fig:AppDecopmositionModel}). In the following, we further explain the other conditions, impact on QAs, and constraints for each pattern.

\begin{table}
    \caption{Application decomposition patterns and strategies}
     \label{tab:AppDecPat}
    \begin{tabularx}{\columnwidth}{|p{3.5cm}|X|}
        \hline
        \textbf{Name} & \textbf{Summary}\\\hline
        Decomposed by subdomains \cite{richardson2018microservices} \cite{AWS} & Define services corresponding to Domain-Driven Design (DDD) subdomains.\\\hline
        Decomposed by business capabilities \cite{richardson2018microservices} \cite{AWS} & Define services corresponding to business capabilities.\\\hline
        Service per team \cite{richardson2018microservices} \cite{AWS} & Break down the application into microservices that individual teams can manage.\\\hline
        Decomposed by transactions \cite{AWS} & An application typically needs to call multiple microservices to complete one business transaction. To avoid latency issues, services can be defined based on business transactions.\\\hline
        Scenario analysis \cite{tusjunt2018refactoring} & Identify the business capabilities by analyzing the nouns and verbs from given business scenarios.\\\hline
        Graph-based approach \cite{kamimura2018extracting} & Identify microservices from the source code of existing monolithic applications by graph clustering and visualization techniques.\\\hline
        Data Flow-Driven (DFD) approach \cite{li2019dataflow} & Follow a top-down approach in which data flow diagrams contain the business requirements that are later partitioned through a formal algebra algorithm for identifying microservices.\\\hline
    \end{tabularx}
\end{table}

 To increase \textit{Flexibility}, \textit{Granularity}, \textit{Reliability}, \textit{Reusability}, \textit{Security}, \textit{Functional suitability}, and \textit {Portability}, \textbf{Decomposed by subdomains} pattern can be used. This pattern guides practitioners in defining each microservice responsibility, boundaries, and relationships with other microservices. To successfully implement this pattern, practitioners need to understand the overall business (see Fig. \ref{fig:AppDecopmositionModel}). In contrast, if microservices need to be defined with respect to business capabilities,  \textbf{Decomposed by business capabilities} pattern can be used. Normally, business capabilities are organized into a multi-level hierarchy and generate business value. This pattern improves the \textit{Granularity}, \textit{Performance}, and \textit{Security} of microservices if the business capabilities are identified by understanding the client organization’s structure, purposes, and business processes. However, this pattern reduces \textit{Flexibility} as the application design is tightly coupled with the business model. Another option that we can use for decomposing applications is \textbf{Service per team} pattern. This pattern enables practitioners to break applications into microservices that individual teams can manage. It also complements \textbf{Decomposed by subdomains} and \textbf{Decomposed by business capabilities} patterns. A constraint of \textbf{Service per team} pattern is that only one small team (e.g., 5-9 people) owns one microservice, meaning that each team independently develops, tests, deploys, and scales individual microservice. The teams also interact with other teams to negotiate APIs. \textbf{Service per team} pattern increases \textit{Availability}, \textit{Scalability}, \textit{Cohesion}, \textit{Deployment}, and \textit{Performance}, and \textit{Maintainability}. If the project is large and needs to hire more people, \textbf{Service per team} pattern negatively impacts the development cost of microservices. 

Another exclusive pattern option in decomposition patterns is \textbf{Decompose by transactions}, in which applications are decomposed based on business transactions. Each business transaction carries one task, and each microservice has the functionalities for several business transactions (e.g., sales, marketing). This pattern allows grouping multiple microservices to avoid latency issues. \textbf{Decompose by transactions} pattern can help to improve \textit{Response time}, \textit{Data consistency}, and \textit{Availability} of microservices. Meanwhile, decomposing applications based on transactions also increases \textit{Execution cost} and \textit{Coupling} of microservices due to multiple functionalities being implemented in one microservice. Another option to decompose an application is \textbf {Scenario-based analysis} which consists of several steps, such as developing scenarios, describing MSA, and evaluating scenarios. During the evaluation process of scenarios, practitioners identify the microservices and interactions between them. This pattern is appropriate if practitioners have enough time to develop and describe the scenarios and MSA, respectively. This strategy can also be used to identify the business capabilities of systems by analyzing the nouns and verbs from given business scenarios. The identified nouns represent the microservices, and the verbs describe the relationship among them. While this strategy increases \textit{Scalability}, \textit{Performance} and \textit{Coupling} could be compromised because of the imprecise boundaries of microservices.

Suppose that the team size is not defined for designing and developing microservices, and we need to identify the microservices from the code of legacy applications. In that case, we can choose \textbf{Graph-based approach}. This approach uses the SArF clustering algorithm to decompose the system for comprehension \cite{kobayashi2013sarf} and the city metaphor techniques for visualizing the system structure \cite{kamimura2018extracting}. \textbf{Graph-based approach} helps to identify microservices from the source code of existing monolithic applications. The use of this approach increases the \textit{Reusability} of the existing code. \textbf{Graph-based approach} also visualizes the extracted microservices and their relationships along with the structure of the whole system. Hence, it also increases the \textit {Understandability} about the MSA design. Finally, if the team size is not defined and applications need to be decomposed by using DFDs, in that case, \textbf{Data flow-driven} approach can be used, which consists of several steps, such as eliciting and analyzing the business requirements for identifying use cases and business logic specifications, creating fine-grained DFDs, identifying the dependencies between processes and datastores, and identifying microservices by clustering processes and related data stores. \textbf{Data flow-driven} approach increases \textit{Availability}, \textit{Scalability}, and \textit{Flexibility}. In contrast, it decreases \textit{Performance} and \textit{Reusability} mainly because of complex DFDs.

\section{Threats to Validity}
\label{sec:threats}
The threats to \textit{construct validity} are related to taking correct operational measures for collecting the data in this study. One potential threat is the inadequate use of the primary constructs of the decision model (i.e., MSA patterns and strategies, QAs, impact of the patterns on QAs). To mitigate this threat, we adopted the following operational measures: (i) we conducted a pilot search to ensure the correctness and appropriateness of the search terms, (ii) we used eight databases (i.e., ACM Digital Library, IEEE Explore, Springer Link, Science Direct, Wiley Online, Engineering Village, Web of Science, and Google Scholar) in software engineering research for retrieving the scientific studies, and (iii) we used Google for searching the grey literature. Additionally, we followed the guidelines \cite{Garousi2019} to review and extract data from the scientific and grey literature.

The threats to \textit{internal validity} represent circumstances that could influence the results obtained from the research. We tried to mitigate this threat through collaborative work between the authors of this work. Regarding the collaborative work, one author proposed the decision model and the rest of the authors contributed to improving the models based on their knowledge and expertise. 

%The potential threats to \textit{external validity} are related to the degree in which the results of a study can be generalized. In this respect, we considered the validation of decision models as a threat. Although the evaluation we conducted is not extensive, the practitioners who participated in the semi-structured interviews were active in the industry and had sufficient expertise and experience in microservices systems. This implies that their evaluation helps to improve the decision models we have proposed.

%The threats to \textit{conclusion validity} affect the ability to reach correct conclusions. In order to mitigate this threat, we defined a research methodology based on the practices and guidelines used in recent studies (e.g., \cite{AR6}, \cite{AR7}, \cite{Garousi2019}) to identify MSA patterns and strategies and to create and evaluate our decision models.% Additionally, to ensure the reliability of our study, we have also made the Replication Package available \cite{replpack}.

\section{Related Work}
\label{sec:relatedWork}

\textit{Decision Models for Architecting Microservices Systems}: The study in \cite{AR3} examines existing literature and provides guidance models for microservices discovery and fault tolerance. The study in \cite{AR4} reports decision guidance models about generating, processing, and managing monitoring data, and disseminating monitoring data to stakeholders in the process automation domain. On the other hand, the study in \cite{AR2} analyzes the strategies and provides guidelines to support architects in selecting suitable frontend architecture(s) for microservices systems. 

\textit{Practitioners' Perspectives and Recommendations for Architecting Microservices Systems}: The research in \cite{AR1} derives a formal architecture decision model containing 325 elements and relations that can help to reduce the (i) efforts needed to understand the architectural decisions and (ii) uncertainty in the design process. An empirical study in \cite{AR5}  interviewed 10 microservices experts to find out the answers to (i) which design areas are relevant for microservices, (ii) how important they are, and (iii) why they are important. 

\textit{Decision Models for Architectural Patterns Selection}: The study \cite{AR5} proposes a decision model that assists developers and architects in selecting appropriate patterns for blockchain-based applications. In a similar study \cite{AR7}, the authors present decision models for cyber-foraging systems that map functional and non-functional requirements to architectural tactics for designing and developing cyber-foraging systems.

\textit{Conclusive Summary}: Our review of the related work suggests that there is a lack of research on decision models that can leverage patterns and strategies as reusable knowledge to address specific design area of microservices systems (i.e., application decomposition). To the best of our knowledge, our proposed decision model that supports decomposing applications into microservices is not covered in any existing studies. This decision model also provides an initial foundation to the research and practice in pattern-based architecting of microservices systems.
\section{Conclusions}
\label{sec:conclusions}
The paper proposes a decision model for selecting patterns and strategies to decompose applications into microservices. The proposed model is constructed by reviewing scientific and grey literature. The decision model provides MSA patterns, strategies, and their impact on QAs for selecting patterns and strategies in decomposing applications into microservices.
In the next step, we aim to (1) propose decision models for other design areas, e.g., microservices security, communication, and service discovery, (2) validate the proposed decision models in an industrial setting, and (3) develop a recommendation system for selecting patterns and strategies for microservices systems.

\section*{Acknowledgments} \label{sec:ack}
This work has been supported by the National Key R\&D Program of China under Grant No. 2018YFB1402800 and the NSFC under Grant No. 62172311.
\bibliographystyle{splncs04}
\bibliography{mybibliography}

\begin{thebibliography}{10}
\providecommand{\url}[1]{\texttt{#1}}
\providecommand{\urlprefix}{URL }
\providecommand{\doi}[1]{https://doi.org/#1}

\bibitem{dragoni2017microservices}
Dragoni, N., Lanese, I., Larsen, S.T., Mazzara, M., Mustafin, R., Safina, L.:
  Microservices: How to make your application scale. In: Proc. of the 11th PSI.
  pp. 95--104. Springer LNCS (2017)

\bibitem{Garousi2019}
Garousi, V., Felderer, M., M{\"a}ntyl{\"a}, M.V.: Guidelines for including grey
  literature and conducting multivocal literature reviews in software
  engineering. Information and Software Technology  \textbf{106},  101--121
  (2019)

\bibitem{AWS}
Hari, O.P.R., Tabby, W., Dmitry, G.: Decomposing monoliths into microservices-
  {AWS} prescriptive (2021), \url{http://alturl.com/5spq4}

\bibitem{AR2}
Harms, H., Rogowski, C., Lo~Iacono, L.: Guidelines for adopting frontend
  architectures and patterns in microservices-based systems. In: Proc. of the
  11th ESEC/FSE. pp. 902--907. ACM (2017)

\bibitem{AR4}
Haselb{\"o}ck, S., Weinreich, R.: Decision guidance models for microservice
  monitoring. In: Proc. of the 14th ICSAW. pp. 54--61. IEEE (2017)

\bibitem{AR3}
Haselb{\"o}ck, S., Weinreich, R., Buchgeher, G.: Decision guidance models for
  microservices: service discovery and fault tolerance. In: Proc. of the 5th
  ECBS. pp. 1--10. ACM (2017)

\bibitem{AR5}
Haselb{\"o}ck, S., Weinreich, R., Buchgeher, G.: An expert interview study on
  areas of microservice design. In: Proc. of the 11th SOCA. pp. 137--144. IEEE
  (2018)

\bibitem{kamimura2018extracting}
Kamimura, M., Yano, K., Hatano, T., Matsuo, A.: Extracting candidates of
  microservices from monolithic application code. In: Proc. of the 25th APSEC.
  pp. 571--580. IEEE (2018)

\bibitem{kobayashi2013sarf}
Kobayashi, K., Kamimura, M., Yano, K., Kato, K., Matsuo, A.: Sarf map:
  Visualizing software architecture from feature and layer viewpoints. In:
  Proc. of the 21st ICPC. pp. 43--52. IEEE (2013)

\bibitem{AR7}
Lewis, G.A., Lago, P., Avgeriou, P.: A decision model for cyber-foraging
  systems. In: Proc. of the 13th WICSA. pp. 51--60. IEEE (2016)

\bibitem{li2019dataflow}
Li, S., Zhang, H., Jia, Z., Li, Z., Zhang, C., Li, J., Gao, Q., Ge, J., Shan,
  Z.: A dataflow-driven approach to identifying microservices from monolithic
  applications. Journal of Systems and Software  \textbf{157},  110380 (2019)

\bibitem{newman2020building}
Newman, S.: Building Microservices: Designing Fine-Grained Systems. O'Reilly
  Media, Inc., second edn. (2020)

\bibitem{AR1}
Ntentos, E., Zdun, U., Plakidas, K., Schall, D., Li, F., Meixner, S.:
  Supporting architectural decision making on data management in microservice
  architectures. In: Proc. of the 13th ECSA. pp. 20--36. Springer LNCS (2019)

\bibitem{richardson2018microservices}
Richardson, C.: Microservices Patterns: With Examples in Java. Manning (2018)

\bibitem{taibi2019monolithic}
Taibi, D., Auer, F., Lenarduzzi, V., Felderer, M.: From monolithic systems to
  microservices: An assessment framework. Information and Software Technology
  \textbf{137},  106600 (2021)

\bibitem{tusjunt2018refactoring}
Tusjunt, M., Vatanawood, W.: Refactoring orchestrated web services into
  microservices using decomposition pattern. In: Proc. of the 4th ICCC. pp.
  609--613. IEEE (2018)

\bibitem{waseemMSAdevops}
Waseem, M., Liang, P., Shahin, M.: A systematic mapping study on microservices
  architecture in {DevOps}. Journal of Systems and Software  \textbf{170},
  110798 (2020)

\bibitem{waseem2021nature}
Waseem, M., Liang, P., Shahin, M., Ahmad, A., Nassab, A.R.: On the nature of
  issues in five open source microservices systems: An empirical study. In:
  Proc. of the 25th EASE. pp. 201--210. ACM (2021)

\bibitem{waseemMSAdesign}
Waseem, M., Liang, P., Shahin, M., Di~Salle, A., Gast\'{o}n, M.: Design,
  monitoring, and testing of microservices systems: The practitioners'
  perspective. Journal of Systems and Software  \textbf{182},  111061 (2021)

\end{thebibliography}
\end{document}